\documentstyle[twocolumn]{jpsj}

\def\runtitle{Magnetic Susceptibility for CaV$_4$O$_9$}
\def\runauthor{K. {\sc Takano} and K. {\sc Sano}}

\title{Magnetic Susceptibility for CaV$_4$O$_9$ }

\author
{ 
Ken'ichi {\sc Takano}
and Kazuhiro {\sc Sano}$^{1}$
}

\inst
{
Laboratory of Theoretical Condensed Matter Physics and \\ 
Research Center for Advanced Photon Technology, \\
Toyota Technological Institute, Nagoya 468-8511 \\
$^1$Department of Physics Engineering, \\
Mie University, Tsu, Mie 514-8507 
}

\recdate{\hskip 5cm}

\abst
{
      We examine experimental magnetic susceptibility 
$\chi^{\rm tot}(T)$ for CaV$_4$O$_9$ by fitting with fitting function 
$\alpha \chi^{\rm mag}(T) + c$. 
     The function $\chi^{\rm mag}(T)$ is a power series of $1/T$ and 
the lowest order term is fixed as $C/T$, where $C$ is the Curie constant 
as determined by the experimental $g$-value ($g$=1.96). 
      Fitting parameters are $\alpha$, $c$ and expansion coefficients 
except for the first one in $\chi^{\rm mag}(T)$. 
     We determine $\alpha$ and $c$ as $\alpha \simeq$ 0.73 and 
$c\simeq$ 0 for an experimental sample. 
     We interpret $\alpha$ as the volume fraction of CaV$_4$O$_9$ 
in the sample and $\chi^{\rm mag}(T)$ as the susceptibility for 
the pure CaV$_4$O$_9$. 
      The result of $\alpha \ne 1$ means that the sample includes 
nonmagnetic components. 
      This interpretation consists with the result of a perturbation theory 
and a neutron scattering experiment. 
}

\kword
{
CaV$_4$O$_9$, magnetic susceptibility, volume fraction, 
high temperature expansion, spin gap, two-dimensional Heisenberg model 
}

\begin{document}
\sloppy
\maketitle


\section{Introduction}

      A few years ago, Taniguchi et al.~\cite{Taniguchi1} measured the 
magnetic susceptibility of layered material CaV$_4$O$_9$ and found 
a spin gap of 107 K by analyzing its temperature dependence. 
      Since then CaV$_4$O$_9$ has been investigated as an interesting 
example of two-dimensional spin systems with spin gap. 
      There is much theoretical effort to understand the spin gap 
by starting from Heisenberg models.~\cite{Katoh1,Ueda,Sano,Troyer,Albrecht,Miyazaki,Gelfand,Fukumoto1,Fukumoto2,Sachdev,White,Starykh,Albrecht2,Takano1,Pickett,Katoh2} 
      However it was not successful to consistently explain various 
experiments. 
      In particular it was very difficult to find the values of exchange 
parameters to reproduce the experimental magnetic susceptibility 
and the $g$-value measured by ESR.~\cite{Taniguchi2} 

\begin{figure}[b]
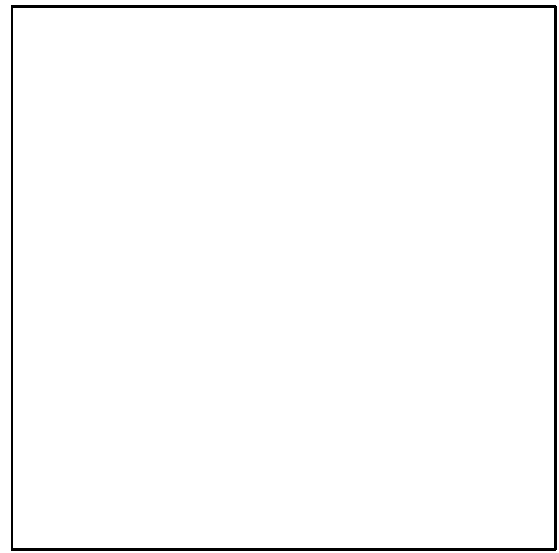

\figureheight{7cm}
\caption{
      Lattice structure for vanadium spins in a layer of 
CaV$_4$O$_9$.  
      The Heisenberg model examined in text includes 
4 dependent exchange parameters $J_e$ (bold solid line), 
$J_e'$ (solid line), $J_c$ (bold dashed line) and 
$J_c'$ (dashed line).}
\label{lattice}
\end{figure}

      From the lattice structure,~\cite{Bouloux} the system is expected 
to be described by the two-dimensional Heisenberg model: 
\begin{equation}
\label{Ham}
      H = \sum_{<i,j>} J_{ij} {\mib S_{i}} \cdot {\mib S_{j}} ,
\end{equation}
where ${\mib S_{i}}$ is the spin at vanadium site $i$. 
      The exchange parameter $J_{ij}$ is nonzero if the pair $<i, j>$ 
corresponds to a link indicated in Fig.~\ref{lattice}: i.~e. $J_{ij}$ is
$J_e$ for an edge-sharing plaquette link, 
$J_e'$ for an edge-sharing dimer link, 
$J_c$ for an corner-sharing plaquette link and 
$J_c'$ for an corner-sharing dimer link. 
      Almost models which have been considered are included in this model 
as special cases. 
%

      We summarize important experimental results as follows: 
\begin{itemize} 
\item[(a)] Temperature dependence of magnetic susceptibility shows 
that the material has a spin excitation gap and its value is about 
110 K.~\cite{Taniguchi1,Isobe} 
      Measurements of the NMR relaxation time confirms 
this value.~\cite{Taniguchi1,Ohama} 
\item[(b)] ESR measurement precisely determined the $g$-value as 
$g$ = 1.96, which is rather close to 2.~\cite{Taniguchi2} 
      The Curie constant is then $C$ = 0.003713 emu/g. 
\item[(c)] Neutron scattering experiment is performed to give a 
dispersion relation for spin excitations.~\cite{Kodama} 
      The result shows a spin gap at momentum $(0, 0)$ and its 
value consists approximately with the values of other experiments. 
      Also the analysis of the scattering intensity directly suggests 
$J_c {>}_{\!\!\!\!\!\!\strut{\sim} } J_e$. 
\end{itemize}

      Theoretically the consistency between the susceptibility and 
the $g$-value is the issue. 
      Before experiments (b) and (c) appeared, Gelfand et al.~\cite{Gelfand} 
estimated the exchange parameters as $J_e \simeq J_e' \simeq$ 190 K 
by assuming $J_c \simeq J_c' \simeq J_e/2$ by using various expansions. 
      In the estimation, they determined the $g$-value as $g$ = 1.77, 
which is different from $g$ = 1.96 in experiment (b). 
      The values for the exchange parameters produce the minimum 
at momentum ($\pi$, $\pi$) in the dispersion relation against (0, 0) in 
experiment (c). 
      In a previous article,~\cite{Sano} we estimated exchange parameters 
as $J_e (\equiv J'_e) \simeq$ 610 K and $J_c (\equiv J'_c) \simeq$ 150 K 
with the assumption of $g = 2$. 
      The estimation comes from the determination of the lowest-order 
coefficient in the high temperature expansion of the susceptibility. 
      To confirm and refine this result we calculated the high-temperature 
expansion to the third order and carried out the fitting to the experimental 
susceptibility data with keeping the measured $g$-value 
($g$ = 1.96).~\cite{Takano1}
      However there is no solution satisfying the obtained set of equations 
to determine the exchange parameters. 
      Katoh and Imada examined various possibilities by taking account of 
plural atomic levels. 
      However their theory includes discrepancy in the compatible 
explanation of the susceptibility and the $g$-value in experiments. 
      These theories shed light on some aspects of the susceptibility of 
this material, although no one succeeded to give a consistent explanation 
of all the experimental results. 

      In this paper, we overcome this difficulty by taking account of 
the possibility that the experimental sample is a mixture of CaV$_4$O$_9$ 
and other nonmagnetic materials. 
      We only assume that the susceptibility can be generally expanded 
in a power series of $1/T$. 
      From the coefficient of $1/T$ in the expansion formula, 
we estimate the volume fraction $\alpha$ of the CaV$_4$O$_9$ component 
in the sample. 
      We further show that the value of $\alpha$ is consistent with 
that obtained by exchange parameters determined by a neutron scattering 
experiment and the third order perturbation calculation.


\section{Determination of volume fraction}

\begin{figure}[t]
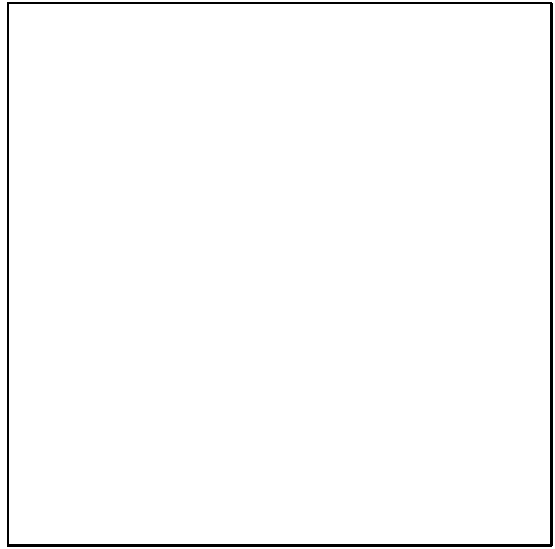

\figureheight{7cm}
\caption{
Experimental susceptibilities of Taniguchi et al.~\cite{Taniguchi1} 
and of Isobe and Ueda~\cite{Isobe}. 
The result of 1.17 times the former is also shown. 
} 
\label{two_experiments}
\end{figure}

      To solve the discrepancy we consider that the susceptibility for the 
sample is smaller than the true value by a constant factor. 
      This is possible if the sample is a mixture and includes nonmagnetic 
components. 
      The constant factor is then the volume fraction of the magnetic 
component. 
      We have two sets of data for susceptibility by Taniguchi 
et al.~\cite{Taniguchi1} and by Isobe and Ueda.~\cite{Isobe} 
      They are shown in Fig.~{\ref{two_experiments}}. 
      Since the latter set is definite for the temperature region under 
300 K, we compared the two sets below 300 K. 
       As a result, we found that the latter is precisely 1.17 times lager 
than the former over all temperatures below 300 K. 
      Hence we infer that the experimental susceptibility includes 
a sample dependent factor corresponding to the amount of nonmagnetic 
components. 
%

      In general, the susceptibility is expanded in a power series of $1/T$ 
and the high temperature behavior is described by the expansion formula. 
      This is true for even the present case where the sample may be 
a mixture. 
      We make a fitting to the experimental data $\chi^{\rm tot}(T)$ by 
a polynomial of sufficiently large order to determine the expansion 
coefficients. 
      To make the expansion coefficients dimensionless we take the 
expansion parameter as $x=T_0/T$ with arbitrary constant $T_0$ of 
the dimension of temperature. 
      We choose it as $T_0$ = 100 K without spoiling the generality. 
      The polynomial for fitting function is accordingly written as 
\begin{equation} 
     f(x) = \frac{C}{T_0} \sum_{m=0}^{L} a_m x^m . 
\label{fx} 
\end{equation}
      Here the expansion coefficients $a_m$'s are fitting parameters. 
      The order $L$ of the polynomial is chosen to be sufficiently large 
so that $a_m$'s do not change largely when $L$ changes by a few integers. 

\begin{figure}[t]
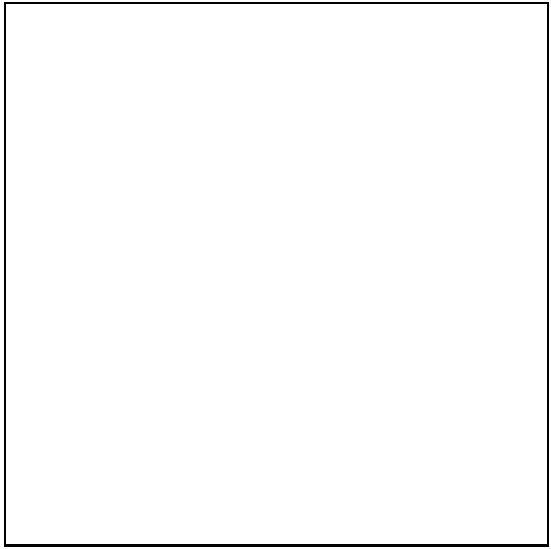

\figureheight{7cm}
\caption{
The fitting function for the optimal fitting to the experimental 
susceptibility. 
Shadowed circles are experimental data $\chi^{\rm tot}$,~\cite{Taniguchi1} 
which seems to form a line by overlap in a high temperature region. 
}
\label{fit}
\end{figure}

      We carried out the fitting for each $L$ less than 19 and obtained 
the values of expansion coefficients. 
      Results for $L$ = 12 to 19 are shown in Table \ref{TableCoef}. 
      Averages over $L$ for 12 to 19 are also shown in the last low. 
      Viewing this table we see that the values for $a_0$ and $a_1$ have 
only small fluctuation and are relatively reliable. 
      In contrast the values for $a_4$ seem to largely fluctuate and 
include an amount of fitting error. 
      The fitted curve is shown in Fig.~\ref{fit} together with 
$\chi^{\rm tot}(T)$. 
%
\begin{table}[t] 
      \caption{Expansion coefficients for susceptibility}
      \label{TableCoef}
\begin{center}
\begin{tabular}{cccccc}
\hline
$L$   & $a_0$  & $a_1$  & $a_2$  & $a_3$  & $a_4$ \\ 
\hline
12  &  0.00117	& 0.72014	& -1.2709	& 1.4170	&-1.0914 \\
13  &  0.00188 	& 0.71095	& -1.2276	& 1.3167	&-0.9596 \\
14  &  0.00039	& 0.73013	& -1.3176	& 1.5224	&-1.2233 \\
15  &  0.00092	& 0.72310	& -1.2830	& 1.4379	&-1.1060 \\
16  & -0.00106	& 0.74988	& -1.4173	& 1.7711	&-1.5742 \\
17  & -0.00048	& 0.74183	& -1.3752	& 1.6604	&-1.4074 \\
18  &  0.00001	& 0.73513	& -1.3409	& 1.5733	&-1.2814 \\
19  &  0.00043 	& 0.72941	& -1.3123	& 1.5028	&-1.1832 \\ 
\hline
Av  &  0.00041	& 0.73007	& -1.3181	& 1.5252	&-1.2283 \\ 
\hline
\end{tabular}
\end{center}
\end{table}
%
%

      The polynomial (\ref{fx}) is rewritten as 
\begin{equation} 
     f(x) = \frac{C}{T_0} [ a_0 + a_1 x (1 + a'_2 x + a'_3 x^2 + \cdots ) ]
\label{fxr} 
\end{equation}
with $a'_i = a_i/a_1$ for $i \ge 2$. 
      Correspondingly we analyze the experimental susceptibility 
$\chi^{\rm tot}(T)$ in the following form: 
\begin{equation}
\label{suscept_tot}
      \chi^{\rm tot}(T) = \gamma \frac{C}{T_0} + \alpha \chi^{\rm mag}(T) , 
\end{equation}
where $\gamma = a_0$ and $\alpha = a_1$. 
      Then $\chi^{\rm mag}(T)$ is approximately represented as 
\begin{equation}
\label{suscpt_mag}
      \chi^{\rm mag}(T)  \simeq
      \frac{C}{T}(1 + A_1 x + A_1 x^2 + \cdots )
\end{equation}
with $A_i = a'_{i+1}$. 
      Hence we estimate the values of $\gamma$ and $\alpha$ as 
\begin{equation} 
      \gamma \simeq 0.00, \quad  \alpha \simeq 0.73. 
\label{gamma-alpha} 
\end{equation}

      For an ideal localized spin system, no constant term appears in the 
magnetic susceptibility. 
      The constant term of $\gamma$ in eq.~(\ref{suscept_tot}) 
comes from other than the electronic spins if it exists. 
      Candidates for the contribution are the Van Vleck term and the diamagnetic term. 
      However these terms have been already subtracted in the present data 
of eq.~(\ref{suscept_tot}).~\cite{Taniguchi1} 
      The present result, $\gamma = 0$, confirms that the original 
subtraction is precise. 
      We note that the present method uses only the general property that 
the susceptibility is expanded in power series of $1/T$. 
      In contrast the method for the original subtraction explicitly uses 
the facts that the system is a spin system and is two-dimensional. 

      If the sample consists only of CaV$_4$O$_9$, the observed 
susceptibility must be $\chi^{\rm mag}(T)$ instead of $\chi^{\rm tot}(T)$. 
      This is because the Curie constant $C$ is fixed as $C$ = 0.003713 
emu/g due to experiment (c) and the constant $\alpha$ must be unity. 
      The result of $\alpha < 1$ means that the sample includes 
nonmagnetic materials, which do not have spin degree of freedom. 
      Precisely the sample is a mixture consisting of 73 \% magnetic and 
27 \% nonmagnetic materials. 
      Thus $\alpha$ is the volume fraction of the magnetic part in 
the whole sample. 
      We infer that the nonmagnetic part is due to incomplete chemical 
reactions and hence CaO and VO$_2$ are candidates. 
      We believe that this is the reason why the effective $g$-value in 
some past theories are much smaller than the observed value by 
the ESR measurement.


\section{Consistency with the Perturbation Theory} 

\begin{figure}[t]
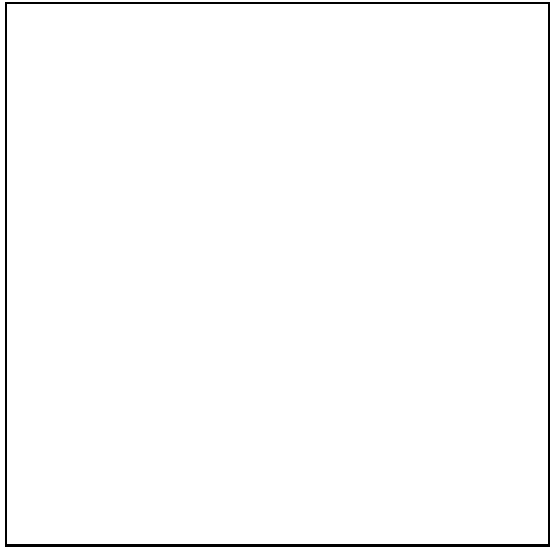

\figureheight{7cm}
\caption{
      Susceptibility obtained by the numerical diagonalization of the 
Hamiltonian (\ref{Ham}). 
      Dash-dotted (dashed) line is for $N$ = 8 ($N$ = 16) lattice. 
      The solid line is for 0.7 times the large $N$ limit of these 
numerical results. 
      Experimental data are also shown. 
      The values of the exchange parameters in the Hamiltonian are 
what Fukumoto and Oguchi~\cite{Fukumoto2} determined. 
}
\label{diagonal}
\end{figure}

      As mentioned in experiment (c), Kodama et al. obtained the 
dispersion relation of a spin excitation for CaV$_4$O$_9$ by neutron 
scattering.~\cite{Kodama} 
      They also theoretically calculated a dispersion formula by 
a perturbation of the second order for the Hamiltonian (\ref{Ham}). 
      By fitting the formula to the experimental dispersion, 
they determined the exchange parameters. 
      The resultant values approximately reproduced the experimental 
dispersion relation. 
      Recently, Fukumoto and Oguchi~\cite{Fukumoto2} recalculated and 
corrected the second order perturbation formula of the dispersion relation. 
      They further extended the perturbation formula up to the third order. 
      In the third order perturbation, their result is given as 
$J_c=162$ K, $J_c=20$ K and $J_e'=J_e=79$ K. 
      Using the third-order values for the exchange parameters, 
we numerically diagonalized the Hamiltonian (\ref{Ham}) and 
obtained energy levels. 
      The diagonalization was done for lattices with $N$ = 8 and 16 
($N$: the number of lattice sites) under the periodic boundary condition. 
      We then calculated the magnetic susceptibility $\chi^{\rm num}(T)$ 
by the energy levels. 
      The results are shown in Fig.~\ref{diagonal}. 
      The curve for 8 sites is close to that for 16 sites. 
      We extrapolate the results to the infinite system size 
by assuming the system size dependence $1/N$. 
      We plotted the values of 0.7 times the extrapolated data in 
Fig.~\ref{diagonal}. 
%
      The resultant curve is close to the experimental susceptibility and 
shows that the volume fraction is given as $\alpha' \sim$ 0.7. 
      This result consists with the interpretation obtained from the 
analysis of the magnetic susceptibility. 

      The difference between $\alpha$ and $\alpha'$ is not serious 
since some errors may be included in the processes determining them. 
      There is a possible error for $\alpha $ in the fitting of a polynomial 
(\ref{fx}) to the experimental data $\chi^{\rm tot}(T)$.~\cite{error} 
      Further there is a possible error in $\alpha'$ accompanied with the 
truncation of the perturbation series to the third order.


\section{Summary}

      We examined the experimentally obtained magnetic 
susceptibility $\chi^{\rm tot}(T)$ for CaV$_4$O$_9$. 
      By using the fact that the susceptibility is generally expanded in 
the power series of $1/T$, we showed that the sample includes 
nonmagnetic components. 
      We estimated the volume fraction of CaV$_4$O$_9$ as 
$\alpha \simeq$ 0.73. 
      The existence of nonmagnetic components enable us to consistently 
understand all the experimental results: the magnetic susceptibility, 
the $g$-value by ESR and the dispersion relation by neutron scattering.


\section*{Acknowledgments}

      This work is partially supported by the Grant-in-Aid for 
Scientific Research from the Ministry of Education, Science 
and Culture, Japan.



\begin{thebibliography}{99}
\bibitem{error} 
      To estimate the accuracy of the fitting method in \S 4, we applied 
the same fitting to the data of the numerical diagonalization ($N = 16$) 
instead of the real experimental data. 
      We found that approximately 2 \% error is involved in the estimated 
coefficient for the first order term, corresponding to the volume fraction.
\end{thebibliography}
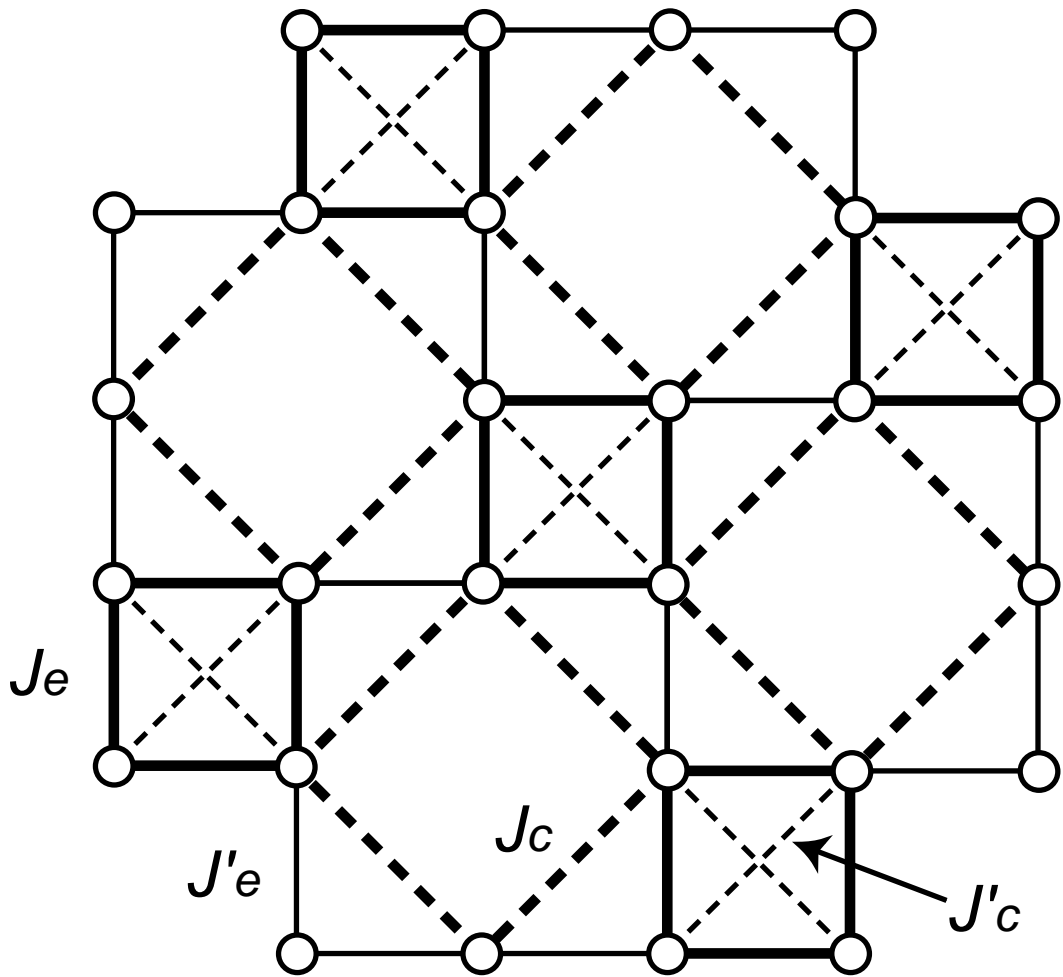
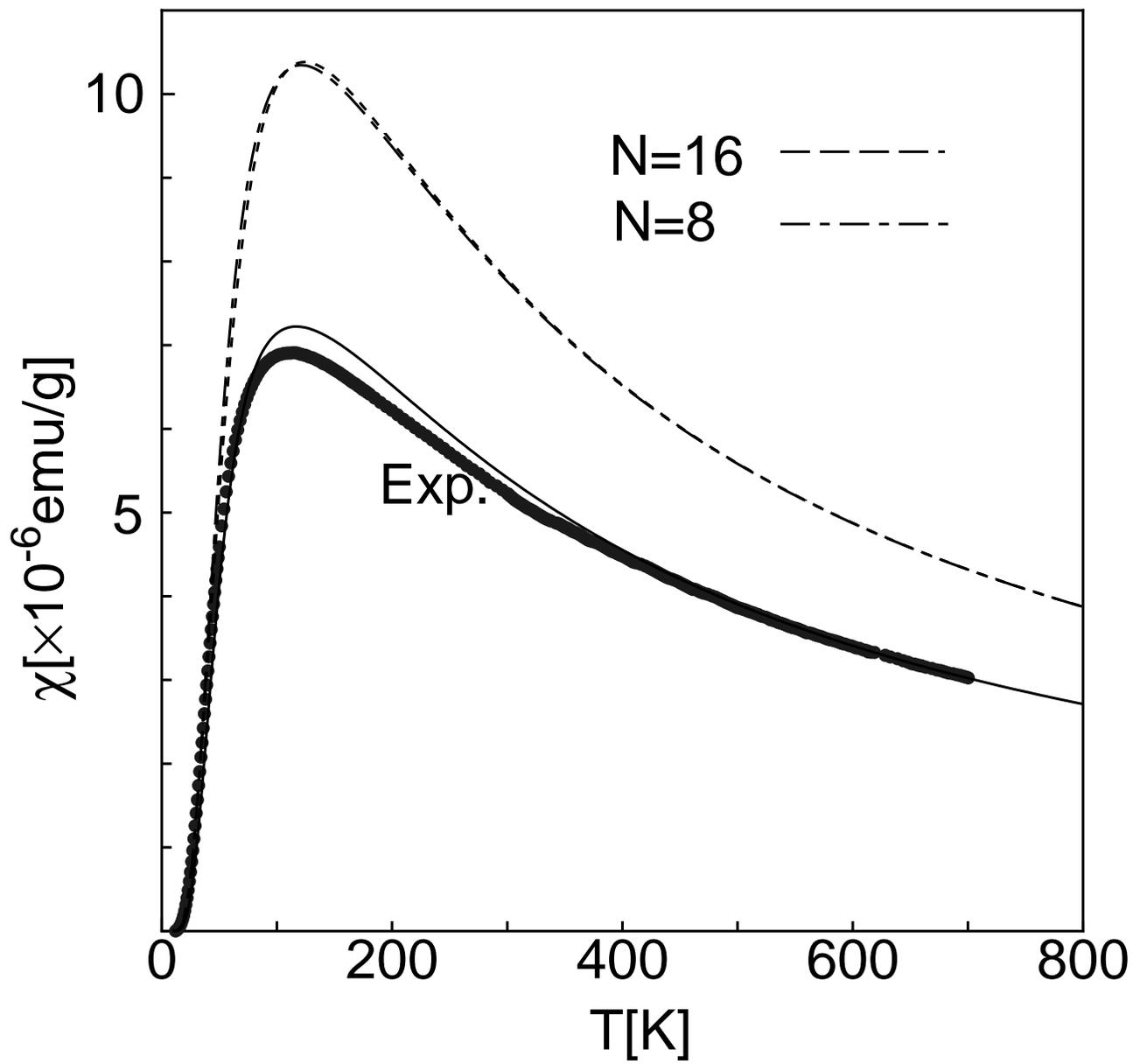
\end{document}